\documentclass[prb,aps,amsmath,amssymb,floatfix,twocolumn,superscriptaddress,groupedaddress]{revtex4-2}
\usepackage{amsmath,bm,graphicx,braket,amssymb,amsfonts}
\usepackage{float}
\usepackage{array, multirow}
\usepackage[utf8]{inputenc}

\usepackage{color}
\usepackage[normalem]{ulem}
\usepackage{subcaption}
\usepackage[T1]{fontenc}
\usepackage{hyperref}
\hypersetup{colorlinks,
	linkcolor=blue, 
	urlcolor=blue,
	citecolor=blue,
}

\begin{document}
	\title{\textbf{Performance Parameters of Infra-red and Visible-active MXene Photocatalysts for Water Splitting}}
	\author{Swati Shaw}
	\email{swatishaw@iitg.ac.in}
	\author{Subhradip Ghosh}
	\email{subhra@iitg.ac.in}
	\affiliation{Department of Physics, Indian Institute of Technology Guwahati, Guwahati-781039, Assam, India.}
	%\date{\today}
	\begin{abstract}
		Water splitting reactions through photocatalysis is an efficient and sustainable technique for the generation of green energy. The photocatalyst's ability to effect simultaneous generation of hydrogen and oxygen, along with efficiency in utilisation of charged carriers, conversion of solar energy to hydrogen, fast migration, and low recombination rates of carriers, are the parameters to decide its suitability in water splitting. In literature, comprehensive calculation and analysis of all these performance parameters for a potential photocatalyst are rare. In this work, we have performed first-principles-based computations to find new efficient photocatalysts from the family of Janus MXenes and assessed their performance parameters. Strain engineering has been invoked in search of new materials. Out of 14 studied materials, we find 5 materials: Sc$_{2}$COS, Zr$_{2}$COS, Hf$_{2}$COS, and ZrHfCO$_{2}$ under zero or finite tensile strain and Hf$_{2}$COSe at 6\% tensile strain meeting the requirements of simultaneous reactions to split water. The computations of various efficiency-related parameters demonstrate that Zr$_{2}$COS, Hf$_{2}$COS, and Hf$_{2}$COSe have excellent efficiencies, significantly better than the well-known photocatalysts. The origin of such performances lies in their electronic and optical properties, which are analysed systematically.   
	\end{abstract}
	\maketitle
	\section{Introduction}
	Photocatalytic water splitting is one of the sustainable techniques for utilising solar energy to produce hydrogen. Economic production of hydrogen from sunlight via photocatalytic splitting of water requires at least 10 \% solar-to-thermal (STH) efficiency \cite{cox2014ten,khaselev2001high}. The STH efficiency depends on light harvesting, separation of carriers, their transportations, hydrogen evolution reactions(HER), and oxygen evolution reactions(OER).To increase the STH efficiency through these steps requires a photocatalyst with an electronic band gap of at least 1.23 eV so that it is greater than or equal to the energy difference between  $H_{2}O \rightarrow O_{2}$(OER) and $H^{+} \rightarrow H_{2}$(HER) reactions. The alignment of the band edges, the optical absorbance, the carrier mobilities, and the binding energy of excitons contribute to this. Assuming 100\% light absorption and quantum efficiency, the theoretically calculated maximum STH efficiency is about 48\% \cite{maeda2010photocatalytic}. However, energy loss due to carrier migration and separation of bound electron-hole pairs is unavoidable. Substantial overpotentials are also required to drive the reactions. Factoring these in, the realistic potential required to drive the photocatalytic process is generally greater than 2.0 V \cite{walter2010solar}. This implies that the band gap of the photocatalyst should be around 2.0 eV, resulting in absorption of only 8\% of the solar spectrum, coming mostly from the ultra-violet (UV) part. This reduces the theoretical STH efficiency to $\sim$ 18\% only\cite{walter2010solar, m2x3}. Many attempts\cite{gai2009design,kudo2009heterogeneous,fu2016two,yang2013roles,osterloh2013inorganic} have been made to improve the STH efficiency that put the choice of material as photocatalyst at the centerstage. 
	
	Enhancement in the STH efficiency is possible if light from the Infrared (IR) region is absorbed as it accounts for nearly 50\% of the solar energy. This requires doing away with the constraint on the electronic band gap of the catalyst material. A proposed mechanism that exploits the presence of an intrinsic electric field along $z$-direction to lift the restriction on band gap \cite{PRL} has led to the prediction of several new IR active photocatalysts from the family of two-dimensional (2D) materials \cite{ramasubramaniam2012large,xu2017,wen2017review}. 2D materials, due to their high surface to volume ratio, a large number of surface reactive sites, excellent optical absorption, high carrier mobility, faster carrier transfer, and low recombination rates, are ideal for catalysing photo-assisted water splitting. These traits, along with the possibility of creating a vertical internal electric field that may enable tapping the IR part of the solar spectrum, project them as ideal compounds to amplify STH efficiency in photocatalytic water splitting. Among them, M$_{2}$X$_{3}$ (M=Al, Ga, In; X=S, Se, Te) turn out to be substantially efficient in solar to hydrogen conversion as several compounds in the series have STH greater than 10\% \cite{m2x3}. For example, IR light-driven photocatalyst In$_{2}$Te$_{3}$, in this series, has a large STH efficiency of 32 \%. Another prominent family of 2D compounds that exhibit both OER and HER simultaneously is the Janus transition metal di-chalcogenides \cite{lu2017janus,ju2020janusMo,ma2018janus,ju2020janus,ramasubramaniam2012large,shen2021janus,fan2020highly,zhao2023janus}. These compounds have higher STH along with higher carrier mobility compared to their parent compounds. The two reactions associated with water splitting can take place on two surfaces of these compounds, leading to fast carrier transport. 
	
	MXenes with chemical formula M$_{n+1}$X$_{n}$T$_{n}$ (M a transition metal, X carbon or nirogen and T a functional group like -O, -F, -S, -Se), a relatively new addition to the family of 2D compounds, has turned out to be a family of compounds where the compositional flexibility can be utilised to obtain a variety of superior functional properties \cite{ma2021ti, zhou2016two,luo2020first,mostafaei2021computational, li2020pressure, he2021switchable,zhang2023theoretical, yang2020identification,kan2020rational,meng2021advances, yu2019surface}. Apart from predicting several M$_{2}$CT$_{2}$ MXenes as potential photocatalysts for water splitting \cite{guo2016mxene,zhang2016computational,cui2023theoretical} by Density Functional Theory(DFT) calculations, possibility to obtain new photocatalysts has been explored by constructing Janus MXenes where either T or M on the two surfaces of $n=1$ MXene are different \cite{zhang2021computational,ozcan2023exploring,wang2022structural,huang2022structural,hong2020double,wong2020high}. In a recent DFT based combinatorial study on 47 Janus M$_{2}$CTT$^{\prime}$ and MM$^{\prime}$CT$_{2}$ MXenes \cite{swati1} four new potential photocatalysts, two IR active and two visible light active, are predicted. Out of these, IR active Zr$_{2}$COS, Hf$_{2}$COS and visible light active Sc$_{2}$COS belong to the sub-family of M$_{2}$COT where M= Sc, Ti, Zr, Hf and T= S, Se. ZrHfCO$_{2}$, the other visible light active photocatalyst belongs to the sub-family of MM$^{\prime}$CO$_{2}$ (M, M$^{\prime}$=Sc, Ti, Zr, Hf). In course of this investigation, it was discovered that Sc$_{2}$COSe from M$_{2}$COT family and TiZrCO$_{2}$, TiHfCO$_{2}$ from MM$^{\prime}$CO$_{2}$ family are semiconductors where due to inappropriate position of valence band maxima (VBM) only HER can take place. Moreover, Ti$_{2}$COS, Ti$_{2}$COSe, Hf$_{2}$COSe, Zr$_{2}$COSe from M$_{2}$COT sub-family and ScTiCO$_{2}$, ScZrCO$_{2}$, ScHfCO$_{2}$ from MM$^{\prime}$CO$_{2}$ group of compounds are semi-metals. Since strain engineering is a well-known cheap way of modulating the band structure of a material, the application of external strain in these compounds may induce semiconducting gap opening in the semi-metals and proper alignments of band edges in the semiconductors, resulting in new catalysts for photo-assisted water splitting. Moreover, the overpotentials associated with the HER in Zr$_{2}$COS and Hf$_{2}$COS are about 0.1 eV only, while those associated with the OER in Sc$_{2}$COS is 0.14 eV. For ZrHfCO$_{2}$, the overpotentials associated with both reactions are about 0.25 eV. The application of strain can be a useful way to enhance these overpotentials. 
	
	None of the works on MXenes available so far considered calculations of the STH efficiency. Another important component to assess the performance of a proposed photocatalyst, the exciton binding energy, is also missing in those works. The coulomb interaction between photo-excited electrons and hole pairs is attractive resulting in excitons, a quasiparticle bound state. The lower the exciton binding energy, the easier the separation of the carriers with opposite charges. This lowers the probability of recombination and thus increases the performance of the photocatalyst. In this work, we address these issues by performing first-principles DFT-based calculations on the above-mentioned 14 Janus MXenes under bi-axial compressive and tensile strains. Apart from the four photocatalysts already predicted, we find Hf$_{2}$COSe under 6\% tensile strain to be a new IR active photocatalyst. We find significant improvement of overpotentials in Zr$_{2}$COS and Hf$_{2}$COS under strain. The calculations of optical spectra, STH efficiency, and exciton binding energies are carried out. The results indicate that Zr$_{2}$COS, Hf$_{2}$COS, and Hf$_{2}$COSe have excellent performance parameters and can be considered as useful IR active photocatalysts for water-splitting. ZrHfCO$_{2}$, on the other hand, has reasonably good performance parameters and can be useful as a visible active photocatalyst.
	
	\section{Computational Details}
	DFT with Projector Augmented Wave-Pseudopotential(PAW-PP) basis \cite{PAW, PAW1}as implemented in Vienna \textit{ab initio} simulation package (VASP)\cite{Vasp, Vasp1} has been used in this work. The Perdew-Burke-Ernzerhof (PBE) parametrisation of Generalized Gradient Approximation (GGA) is used to approximate the exchange-correlation part of the Kohn-Sham Hamiltonian. Plane waves up to 550 eV are considered. The convergence thresholds for Hellmann-Feynman force and total energy are chosen as 10$^{-3}$ eV/\AA and 10$^{-6}$ eV, respectively. A $\Gamma$-centered $k$-point grid of 9$\times$9$\times$1 is used for geometry optimization. A 20 \mbox{\AA} vacuum slab is inserted along the z-direction to minimize the interaction between adjacent layers. The more accurate Heyd-Scuseria-Ernzerhof (HSE06) \cite{HSE1} hybrid functional with dipole correction along  $z$ direction is used to calculate electronic band structures. The elastic constants are obtained using a stress-strain relation with five different distortions of the lattice. 
	
	The dynamical stability of a system is investigated by computing the phonon dispersions. Density Functional Perturbation Theory method \cite{baroni2001phonons} as implemented in PHONOPY code \cite{togo2015first} is used to compute phonon spectra. Phonon calculations are performed in a 4$\times$4$\times$1 supercell with a $\Gamma$ centered 3$\times$3$\times$1 $k$-mesh. \textit{Ab initio} molecular dynamics (AIMD) simulations are done to check the thermal stability of systems. The calculations are done with 3$\times$3$\times$1 supercells at room temperature(300K). Nose-Hoover thermostat \cite{nose1984unified} is used. Simulations are run up to 15 ps with a time step of 2 fs. The optical absorption spectra are obtained by computing frequency-dependent dielectric functions using HSE06 functional where the Brillouin Zone is sampled with a 9$\times$9$\times$1 $k$-grid.
	\begin{figure*}[htb!]
		\centering
		\includegraphics[scale=0.4]{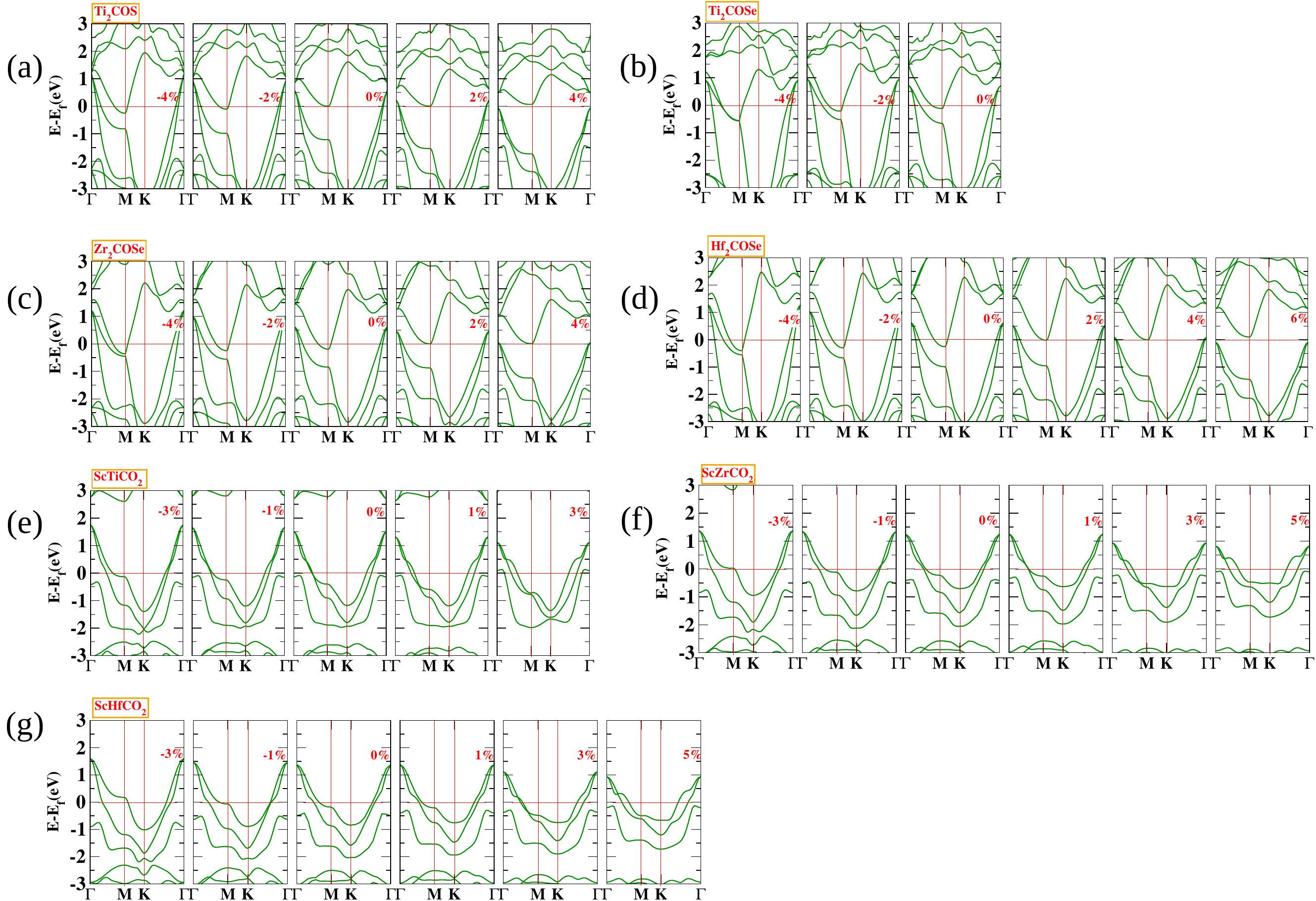}
		\caption{The Electronic Band Structures of a) Ti\textsubscript{2}COS, b) Ti\textsubscript{2}COSe, c) Zr\textsubscript{2}COSe, d) Hf\textsubscript{2}COSe, e) ScTiCO$_2$, f) ScZrCO$_2$ and g) ScHfCO$_2$ under various Biaxial strains. The Fermi level is marked with a horizontal red line. }
		\label{Figure1}
	\end{figure*}
	\section{Results and Discussions}
	As it has been mentioned earlier, 14 Janus MXenes, 8 from M$_{2}$COT family and 6 from MM$^{\prime}$CO$_{2}$ family have been considered in this work. 
	Biaxial strain defined as $\bm{\alpha} = \left(\dfrac{a -a_0}{a_0}\right) \times 100\%$, where $a(a_0)$ is the lattice constant of the strained (unstrained) Mxene monolayers, has been applied on them. $\bm{\alpha}>(<)0$ implies tensile(compressive) strain. $\alpha$ is varied from -4 to 6(-3 to 5)\% for M$_{2}$COT(MM$^{\prime}$CO$_{2}$) compounds.
	\subsection{Mechanical, Dynamical and Thermal Stabilities of the Compounds}
	Before investigating the effects of strain on the band structure and other properties directly associated with the photocatalyst's utility, it is important to make sure the compounds are stable under applied strain. To this end, we first look at the mechanical stability of the MXenes considered. This is done by calculating their elastic moduli at the respective equilibrium volumes. We have computed the three elastic constants $C_{11},C_{12}$ and $C_{66}=\left(\frac{C_{11}-C_{12}}{2} \right)$ and derived Poison's ratio $\nu=\frac{C_{12}}{C_{11}}$ and Young's Modulus $Y=\left( \dfrac{C_{11}^2-C_{12}^2}{C_{11}}\right) $ from them. The results are presented in Table S1, supplementary information. We apply the Born-Huang's criteria $C_{11} > 0, C_{11} > |C_{12}|, C_{66} >0$ to assess mechanical stability. All 14 compounds investigated satisfy the criteria and are, therefore, mechanically stable. 
	
	The dynamical stability is examined by computing phonon dispersion relations of the compounds by varying $\alpha$. The results are shown in Figures S1-S3, supplementary information. Among the MXenes in M$_{2}$COT series, all except Ti$_{2}$COSe are dynamically stable with strain between -4 \% and 4\%. Ti$_{2}$COSe is stable at compressive strain only. On the other hand, Hf$_{2}$COS and Hf$_{2}$COSe have no imaginary modes even upto $\alpha=6 \%$, the highest strain applied. Among the MM$^{\prime}$CO$_{2}$ compounds, only ScZrCO$_{2}$, ScHfCO$_{2}$ and ZrHfCO$_{2}$ have stable phonon modes for the entire range of strain ($\alpha$=-3\% to 5\%). ScTiCO$_{2}$ is dynamically stable upto 3\% strain while TiZrCO$_{2}$ and TiHfCO$_{2}$ maintain their stability only upto a strain of 1\%. The results suggest that the presence of Zr/Hf as one of the transition metals provides a greater range of stability in Janus MXenes. The thermal stability of each one of these MXenes at room temperature is then assessed by looking at the variations in their temperature and free energy with simulation run time. The calculations are done only at the highest value of $\alpha$, where the systems are dynamically stable. The results are shown in Figures S1-S3, supplementary information. We find that all the structures except ScTiCO$_2$ remain intact at this temperature and strain, indicating thermal stability in all of them. ScTiCO$_2$, although it is dynamically stable up to 3\%of tensile strain, the structure is thermally stable up to only 1\% strain. For subsequent calculations, we have considered only the range of $\alpha$ where each one of the compounds is dynamically stable. 
	
	Therefore, -4\% to 4\% strains are applicable on Sc\textsubscript{2}COS, Sc\textsubscript{2}COSe, Zr\textsubscript{2}COS, Zr\textsubscript{2}COSe and Ti\textsubscript{2}COS except Ti\textsubscript{2}COSe, whereas -4\% to 6\% strains on Hf\textsubscript{2}COS \& Hf\textsubscript{2}COSe. In the case of TiZrCO\textsubscript{2} and TiHfCO\textsubscript{2}, we will use up to 1\% strain and for ZrHfCO\textsubscript{2} up to 5\% tensile strain for upcoming calculations.
	\begin{table*}[htb!]
		\small
		\centering
		\caption{Variations in the Electronic Band Gap (E$_g$), Electrostatic Potential Difference ($\Delta\Phi$) between two surfaces, HER Overpotential ($\chi\left(H_2 \right)$) and OER  Overpotential ($\chi\left(O_2 \right)$) with Biaxial strain $\alpha$.}
		\label{Table1}
		\begin{ruledtabular}
			\begin{tabular}{cccccccccccc}
				\bf{Compounds}  & \textbf{$\bm{\alpha}$} & \bf{E$_g$} & \bf{$\Delta\Phi$} & \bf{$\chi\left(H_2 \right)$} & \bf{$\chi\left(O_2 \right)$} & \bf{Compounds}  & \textbf{$\bm{\alpha}$} & \bf{E$_g$} & \bf{$\Delta\Phi$} & \bf{$\chi\left(H_2 \right)$} & \bf{$\chi\left(O_2 \right)$} \\
				&  (\%)  & (eV) & (eV) & (eV) & (eV) &  &(\%)  & (eV) & (eV) & (eV) & (eV)\\
				\hline
				\multirow{5}{*}{Sc\textsubscript{2}COS}&-4&2.57& 0.73&2.06 & 0.02&\multirow{5}{*}{Sc\textsubscript{2}COSe}&-4& 1.62&0.64 &1.86 &-0.83\\
				&-2& 2.64 &1.20 &2.52 &0.09& &-2& 1.90&1.13 &2.48 &-0.68 \\
				&0&2.57 & 1.58&2.78 &0.14 & &0&1.97 &1.52 &2.81 &-0.55\\
				&2&2.44 &1.88 &2.92 &0.18 & &2&1.91 &1.82 &2.94 &-0.44 \\
				&4& 2.28& 2.13&2.99&0.20 & &4&1.82&2.07 &3.00 &-0.35 \\
				\hline
				\multirow{5}{*}{Zr\textsubscript{2}COS}&-2& 0.04&1.54 &-0.31 & 0.66 & \multirow{5}{*}{Hf\textsubscript{2}COS}&-2&0.01 &1.44 &-0.28 &0.50 \\
				&0& 0.36&1.78 &0.05 &0.86& &0&0.35 &1.70 &0.09 &0.73  \\
				&2&0.63 & 1.94&0.35 &0.98 & &2&0.66&1.86 &0.38 &0.91 \\
				&4&0.86 & 2.10&0.62 &1.11 & &4&0.90 &2.02 &0.64 &1.05\\
				& & & & & & & 6 &1.10 &2.16 &0.86 &1.17 \\
				\hline
				Ti\textsubscript{2}COS& 4& 0.12&1.17 &-0.37 &0.44 & Hf\textsubscript{2}COSe&6&0.20 &2.22 &0.92 &0.27 \\
				\hline
				\multirow{4}{*}{TiZrCO\textsubscript{2}}&-3&1.19&0.07 &-0.73 &0.76 & \multirow{6}{*}{ZrHfCO\textsubscript{2}}&-3&1.41 &0.05 &-0.23 &0.47 \\
				
				&-1&1.41 &0.05 &-0.41 &0.64 & &-1& 1.63&0.04 &0.12 &0.32 \\
				
				&0&1.50 &0.04 &-0.27 &0.58 &&0&1.72& 0.03&0.29&0.24 \\
				
				&1&1.59 &0.02 &-0.14 &0.52 & &1& 1.80&0.03 &0.43 &0.17 \\
				\cline{1-6}
				\multirow{4}{*}{TiHfCO\textsubscript{2}}&-3&1.15&0.00 & -0.73&0.65 & &3&1.94&0.02 &0.69 &0.03 \\
				
				&-1 &1.35&0.00 &-0.42 &0.54 & &5&2.04 &0.01 &0.91 &-0.09 \\
				
				&0&1.47 &0.00 &-0.28 &0.52 & & & & & & \\
				
				&1&1.53 &0.00 &-0.16 &0.46  & & & & & & \\
			\end{tabular}
		\end{ruledtabular}
	\end{table*}
	\begin{figure*}[htb!]
		\centering
		\includegraphics[scale=0.183]{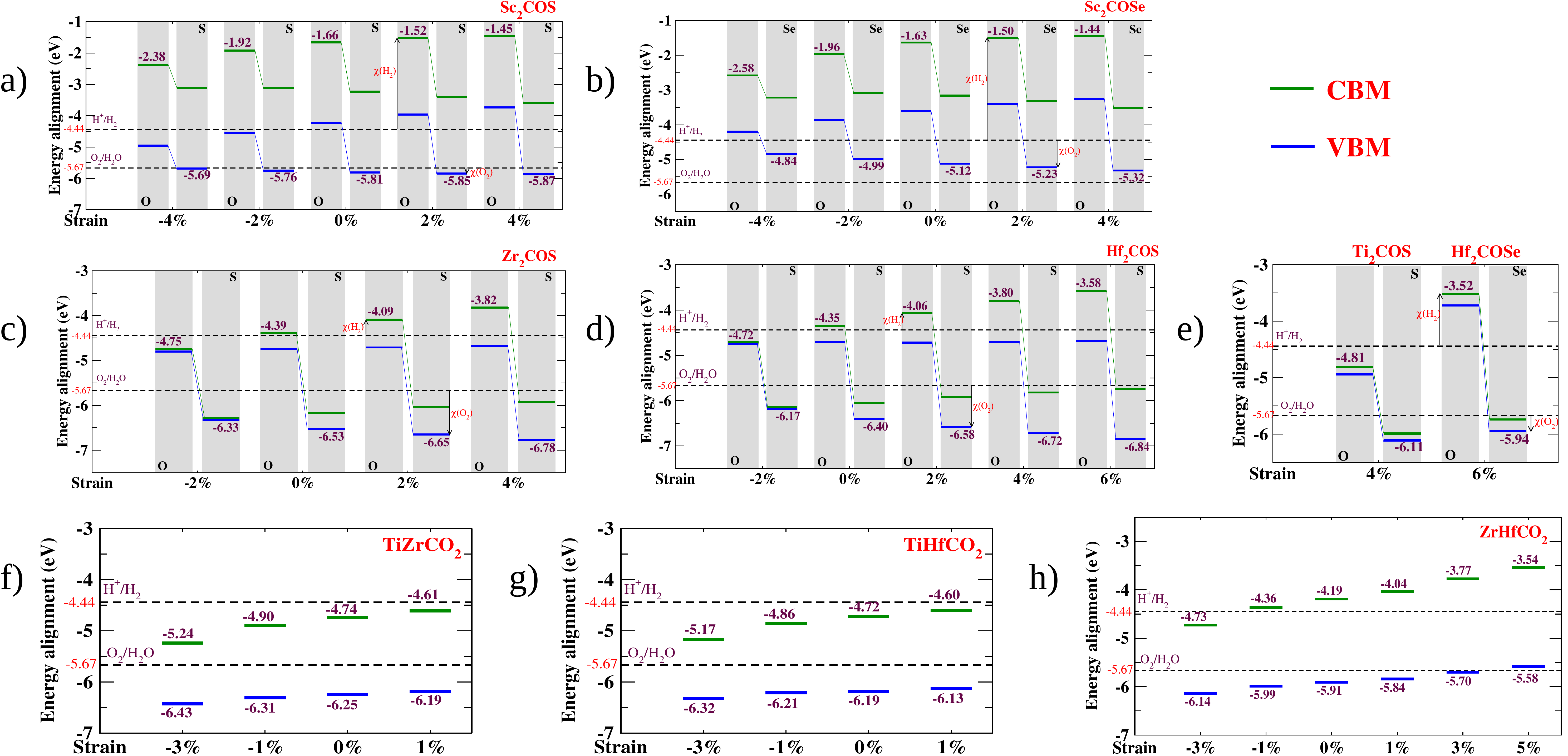}
		\caption{Variations in the Band Edge Positions of a) Sc\textsubscript{2}COS, b) Sc\textsubscript{2}COSe, c) Zr\textsubscript{2}COS d) Hf\textsubscript{2}COS, f) TiZrCO\textsubscript{2}, g) TiHfCO\textsubscript{2} \& h)  ZrHfCO\textsubscript{2} with strain $\alpha$. e) shows the Band Edge Positions of Ti\textsubscript{2}COS at 4\%  \& Hf\textsubscript{2}COSe at 6\% tensile strain. The Overpotentials for M$_{2}$COT MXenes are marked with arrows. The redox (HER and OER) potentials are marked by a dotted black line. Vacuum levels of the both surfaces are shifted to 0 eV.}
		\label{Figure2}
	\end{figure*}
	
	\subsection{Electronic Band Structures and Alignment of Band Edges}
	The motivation behind strain engineering in this work is to modulate the band structures so that a semiconducting gap opens in the Janus MXenes that are semi-metals without strain and the alignments of conduction band minima (CBM) and valence band maxima (VBM) align appropriately in compounds that are semiconductors in their equilibrium volume as well as in the ones that become semiconductors under strain. In Figure \ref{Figure1} (Figure S4, supplementary information), we present the results on electronic band structures as a function of the strain of the Janus compounds that are semi-metals(semiconductors) at equilibrium, that is, with $\alpha=0$. The tensile strain of 4\% (6\%) opens a semiconducting gap of 0.12 (0.2) eV in Ti$_{2}$COS(Hf$_{2}$OSe). The other 5 semi-metals remain metallic even after being subjected to a sizeable strain. In both cases, the semiconducting gaps are indirect with VBM at $\Gamma$ and CBM at M-point. The 7 semiconducting Janus MXenes remain semiconductors under tensile strain. Compressive strain of 4\% induces semiconductor-metal transition in Zr$_{2}$COS and Hf$_{2}$COS only (Figure S4(c),(d), supplementary information). For these 7 compounds, the locations of CBM and VBM in the Brillouin zone do not change with strain; the only change is in the size of the electronic band gap. From Table \ref{Table1}, we find that while band gap $E_{g}$ in MM$^{\prime}$CO$_{2}$ MXenes increases monotonically with $\alpha$, it is not so for the M$_{2}$COT MXenes. The compressive strain drastically decreases $E_{g}$ in Zr$_{2}$COS and Hf$_{2}$COS turning them almost metallic at 2\% strain. This explains their transformation to metal upon further increase in compressive strain. The trend of decrease in $E_{g}$ with an increase in compressive strain is also observed in Sc$_{2}$COSe and Sc$_{2}$COS. Since $E_{g}$ in these compounds are about 2-2.6 eV when no strain is applied, no dramatic change in electronic state is obtained when they are subjected to compressive strain. With tensile strain, the band gaps in  Zr$_{2}$COS and Hf$_{2}$COS increase monotonically. The trend is opposite for the two Sc$_{2}$COT MXenes. Inspite of its increase with $\alpha$, the maximum$E_{g}$ of IR-active Zr$_{2}$COS and Hf$_{2}$COS are about 1.0 eV. This suggests that these two MXenes can continue to be IR-active under tensile strain.   
	
	To understand the trends in the band gap $E_{g}$ with $\alpha$ and to find out how strain modulates the positions of VBM and CBM, in Figure \ref{Figure2}, we show the band edge positions of these nine semiconducting MXenes at different strains. In each case, the alignments of band edges are shown only for the set of $\alpha$ for which the compound is semiconducting. For Sc$_{2}$COT MXenes, both VBM and CBM move to higher energies as strain changes from compressive to tensile. The decrease in $E_{g}$ with $\alpha$ in the tensile region is mostly due to a gradual decrease in the change in the position of CBM. For Sc$_{2}$COSe in particular, larger changes in $E_{g}$ in the compressive region are due to larger changes in the position of the CBM. For other compounds, though both bands move towards higher energies as strain increases from compressive to tensile, the changes in the positions of CBM are about 3 times larger than that of VBM. This explains the trends of $E_{g}$ shown in Table \ref{Table1}. In Reference \cite{swati1}, it was shown that a substantial electric field acts across the surfaces in M$_{2}$COT MXenes, resulting in a sizeable amount of band bending and subsequent re-alignment of VBM and CBM. The resulting potential difference $\Delta\Phi$ between the surfaces lifted the restriction on $E_{g}$ as the band gap is now modified to $(1.23- \Delta\Phi)$ eV. As a consequence, Zr$_{2}$COS, Hf$_{2}$COS(Sc$_{2}$COS) turned out to be potential photocatalysts active in the IR(visible)-region. $\Delta \Phi$ between the two surfaces of M$_{2}$COT MXenes are presented in Table \ref{Table1}. The potential profiles of the compounds are shown in Figure S6, supplementary information. We find that for all compounds, $\Delta \Phi$ increases monotonically with $\alpha$ from compressive to tensile region. The substantial values of $\Delta \Phi$ effect band straddling in Sc$_{2}$COS under compressive strain. Consequently, Sc$_{2}$COS can be considered as a photocatalyst where both OER and HER can take place simultaneously for the entire range of strains considered. In Sc$_{2}$COSe, $\Delta \Phi$ and the subsequent electric field are not enough to push the VBM below the OER potential. As a result, Sc$_{2}$COSe continues to be a catalyst for HER only. Compressive strain in IR-active Zr$_{2}$COS and Hf$_{2}$COS reduce $\Delta \Phi$ significantly destroying the alignment of CBM. Both HER and OER can continue to happen in these two compounds only for tensile strain. A large $\Delta \Phi$ in Hf$_{2}$COSe under 6\% tensile strain generates a large electric field re-aligning the CBM and VBM at positions suitable for OER happening at the Se surface and HER at the O surface. In Ti$_{2}$COS at 4\% tensile strain, $\Delta \Phi$ nearly half that of Hf$_{2}$COSe turns out to be not enough to push the CBM above that of HER potential. Consequently, Ti$_{2}$COS cannot be considered as a photocatalyst for water splitting even with finite strain. Application of strain fails to induce band straddling in TiZrCO$_{2}$ and TiHfCO$_{2}$. A near-zero internal electric field is not enough to supply the necessary energy to push the CBM at energies above HER energy. In ZrHfCO$_{2}$, the appropriate band alignments at zero strain are destroyed when $\alpha$ varies outside the range of [-1\% : 3\%]. Thus, only Sc$_{2}$COS, Zr$_{2}$COS, Hf$_{2}$COS, Hf$_{2}$COSe, ZrHfCO$_{2}$ for a range of strains can be considered for assessment of other parameters associated with photocatalytic efficiencies. It is worth noting that the overpotential $\chi \left(H_2 \right)$ that is calculated as the difference between the CBM and HER energy has improved significantly upon application of tensile strain in Zr$_{2}$COS and Hf$_{2}$COS, a 0.57eV and 0.55eV jump for the former with a 4\% change in $\alpha$. For the later, the change is 0.77 eV when $\alpha$ changes by 6\%. For Sc$_{2}$COS, overpotential $\chi \left(O_2 \right)$, calculated as the difference between the VBM and the OER energy, improves marginally while it degrades considerably in ZrHfCO$_{2}$ as $\alpha$ increases in the tensile region. For the new compound Hf$_{2}$COSe at 6\% tensile strain, both overpotentials are substantial.  
	\begin{figure*}[htb!]
		\centering
		\includegraphics[scale=0.12]{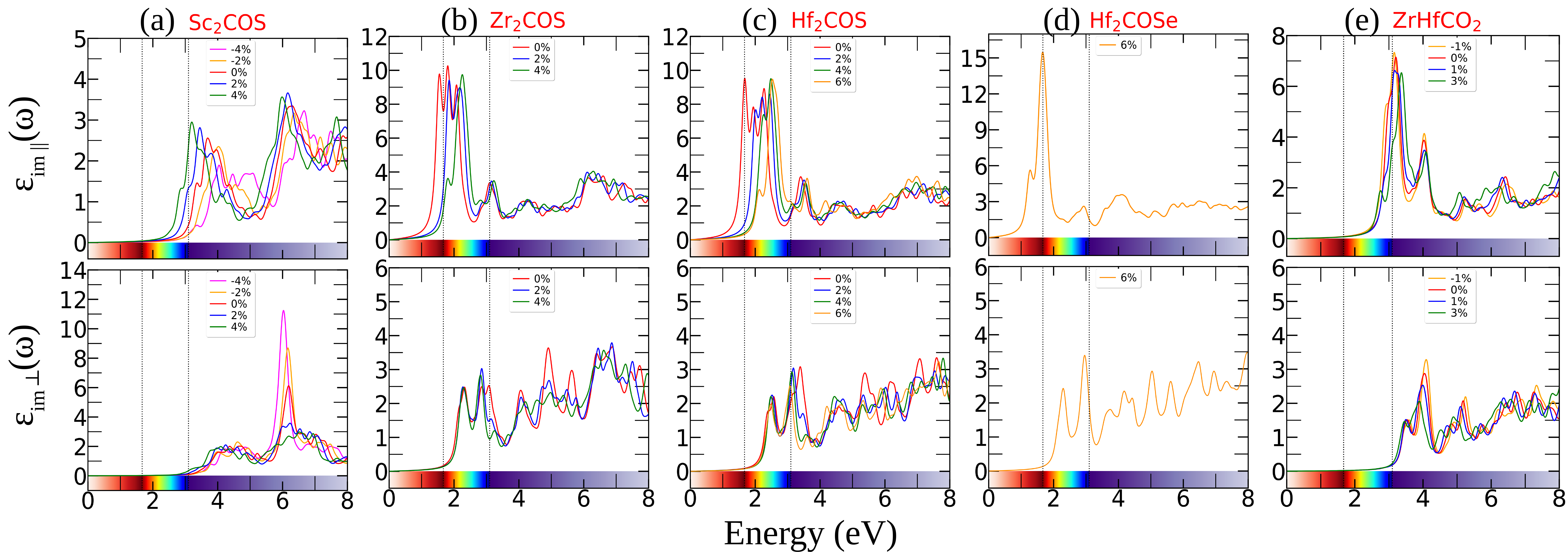}
		\caption{Variations in the Imaginary Part of Dielectric Functions of a) Sc\textsubscript{2}COS, b) Zr\textsubscript{2}COS, c) Hf\textsubscript{2}COS, d) Hf\textsubscript{2}COSe and e) ZrHfCO\textsubscript{2} with biaxial strain $\alpha$. The visible regions are marked by two black dotted lines. The $\parallel$ and $\perp$ symbols express the in-plane and out-of-plane directions, respectively.}
		\label{Figure3}
	\end{figure*}
	\subsection{Optical Properties}
	In order for good STH efficiency, a photocatalyst should have high absorption ability in the infrared and visible part of the solar spectrum. This can be assessed by analysis of complex frequency-dependent dielectric function $\varepsilon(\omega)$ =$\varepsilon_{r}(\omega)$ + $i\varepsilon_{im}(\omega)$. The imaginary part of the dielectric function ($\varepsilon_{im}$) is obtained by summing over the empty states \cite{gajdos2006electronic}:
	\begin{equation*}
	\begin{split}
	\varepsilon^{im}_{\alpha\beta}(\omega)= \dfrac{4\pi^2 e^2}{\Omega}\lim_{q\longrightarrow 0} \dfrac{1}{q^2}\sum_{c,v,k}2\omega_k\delta(\varepsilon_{ck}-\varepsilon_{vk}-\omega)\times\\
	\left\langle u_{ck+e_\alpha q}|u_{vk}\right\rangle\left\langle u_{vk}|u_{ck+e_\beta q}\right\rangle 
	\end{split}
	\end{equation*}
	$e$ is the electronic charge, $\Omega$ is the volume of the supercell, $c$ and $v$ denote the conduction and valance band states, respectively, $\omega$ is the frequency of the incident light. $u_{ck}$ is the cell-periodic part of the orbitals at the point \textbf{k}. The excitation is represented in terms of Dirac $\delta$.
	The real part of the dielectric function can be calculated using the Kramers-Kronig transformation :
	\begin{equation}
	\varepsilon^{r}_{\alpha\beta}(\omega)= 1+\dfrac{2}{\pi}P\int^{\propto}_0 \dfrac{\varepsilon^{im}_{\alpha\beta}(\omega')\omega'}{\omega'^2-\omega^2+i\eta}d\omega'
	\end{equation}
	$P$ stands for the Cauchy principal value of the integral, and $\eta$ is the infinitesimal shift in the complex plane. Using $\varepsilon_{im}$ \& $\varepsilon_{r}$, the absorption coefficient $A(\omega)$ can be calculated as \cite{saha2000structure}:
	\begin{equation}
	A(\omega)=\dfrac{\sqrt{2}\omega}{c}\left\lbrace \left[ \varepsilon_{r}^2(\omega)+\varepsilon_{im}^2(\omega)\right] ^{1/2}-\varepsilon_{r}(\omega)\right\rbrace ^{1/2}
	\label{abs_coeff}
	\end{equation}
	The imaginary part of the dielectric function denotes the absorption of solar energy during the transfer of electrons from VBM to CBM. In Figure \ref{Figure3}, we present the in-plane and out-of-plane components of $\varepsilon_{im}\left(\omega \right)$ of the five Janus MXenes with variation of strain. Due to the strong anisotropy in the Janus MXenes, the in-plane ($\varepsilon_{im\parallel}$) and out-of-plane ($\varepsilon_{im\perp}$) components are quite different. The out-of-plane absorptions in each system happen mainly in the UV part of the solar spectrum. This implies that to get high absorption efficiency in the IR region, the MXene sheets should be held parallel to the light source. 
	
	Upon inspecting the in-plane component $\varepsilon_{im\parallel}$ of Sc$_{2}$COS, we find that in the absence of strain, there are two strong absorption peaks at 3.69 eV and 6.3 eV, both in the UV-region. In the tensile-strain regime, there is a red shift of the absorption spectra due to a gradual decrease in the band gap with an increase in strain. At $\alpha=4\%$, the peak at 3.69 eV shifts to 3.2 eV. An exact opposite trend (blue shift) is observed in the case of out-of-plane component $\varepsilon_{im\perp}$. Variations in the $\varepsilon_{im\parallel}$ of Zr\textsubscript{2}COS and Hf\textsubscript{2}COS with strain have similar characteristics. With no strain, both compounds show the strongest absorption in the IR region. The absorption spectra of Zr$_{2}$COS are characterised by a strong peak at 1.55 eV in the IR region followed by two peaks at 1.81 eV and 2.08 eV, both in the visible region. Here, too, the increase in tensile strain blue shifts the absorption spectra. This implies that zero or low tensile strain in these two compounds would enable them to tap the maximum part of the solar spectrum. The highest absorption in the IR region is observed in the case of  Hf\textsubscript{2}COSe under 6\% tensile strain. The absorption spectra are characterised by a single strong peak at 1.67 eV(Fig.\ref{Figure3}d)). Since ZrHfCO$_{2}$ is predicted to be a visible-active photocatalyst, it is expected that it will absorb mostly from the visible part of the solar energy. In Figure \ref{Figure3}(e), we find that its absorption spectra at zero strain are characterised by a strong peak at around 3.2 eV followed by smaller peaks at higher energy. Application of strain does not induce any significant shift of the spectrum suggesting that its absorption is completely in the visible and UV part of the solar spectrum. 
	The results indicate that strained Hf$_{2}$COSe, Hf$_{2}$COS, and Zr$_{2}$COS under zero or small tensile strain can be photocatalysts with good STH conversion efficiency as their absorptions in the IR region are significant. Sc$_{2}$COS and ZrHfCO$_{2}$ will have STH efficiencies lower than these three as they absorb mostly visible and UV solar light. 
	\begin{table*}[htb!]
		\small
		\caption{Variations in the Energy Conversion Efficiency and Quantum Efficiency of Total ($\eta_{abs}$, $\eta^Q_{abs}$), Ultraviolet ($\eta_{UV}$, $\eta^Q_{UV}$), Visible ($\eta_{VIS}$, $\eta^Q_{VIS}$) and Infrared ($\eta_{IR}$, $\eta^Q_{IR}$)light absorption, Carrier Utilisation ($\eta_{cu}$, $\eta^Q_{cu}$), Solar-to-Hydrogen ($\eta_{STH}$, $\eta^Q_{T}$), Corrected Solar to Hydrogen ($\eta'_{STH}$), with Biaxial strain $\alpha$ for the five Janus compounds.}
		\label{Table2}
		\begin{ruledtabular}
			\begin{tabular}{cccccccccccccccc}
				\textbf{Compounds}&$\bm{\alpha}$&\multicolumn{7}{c}{\textbf{Energy Conversion Efficiencies}}&&\multicolumn{6}{c}{\textbf{Quantum Efficiencies}}\\
				\cline{3-9}\cline{11-16}
				&(\%)&$\bm{\eta_{abs}}$&$\bm{\eta_{cu}}$&$\bm{\eta_{STH}}$&$\bm{\eta'_{STH}}$&$\bm{\eta_{UV}}$&$\bm{\eta_{VIS}}$&$\bm{\eta_{IR}}$&&$\bm{\eta^Q_{abs}}$&$\bm{\eta^Q_{cu}}$&$\bm{\eta^Q_{T}}$&$\bm{\eta^Q_{UV}}$&$\bm{\eta^Q_{VIS}}$&$\bm{\eta^Q_{IR}}$\\
				&&(\%)&(\%)&(\%)&(\%)&(\%)&(\%)&(\%)&&(\%)&(\%)&(\%)&(\%)&(\%)&(\%)\\
				\hline
				\multirow{5}{*}{Sc\textsubscript{2}COS}&-4&13.75&7.86&1.08&1.05&31.03&0.00&0.00&&6.66&18.65&1.24&85.42&0.00&0.00\\
				
				&-2&11.80&9.17&1.08&1.03&31.03&0.00&0.00&&5.60&22.19&1.24&85.42&0.00&0.00\\
				
				&0&13.75&11.78&1.62&1.51&36.32&0.70&0.00&&6.66&27.96&1.86&100.00&1.22&0.00\\
				
				&2&17.52&15.28&2.68&2.40&36.32&2.90&0.00&&8.78&35.02&3.07&100.00&5.08&0.00\\
				
				&4&22.87&19.26&4.41&3.73&36.32&6.51&0.00&&11.99&42.21&5.06&100.00&11.39&0.00\\
				
				\hline
				\multirow{3}{*}{Zr\textsubscript{2}COS}&0&99.44&84.21&83.74&37.55&36.32&57.14&100.00&&97.63&98.50&96.17&100.00&100.00&94.28\\
				
				&2&95.68&80.01&76.55&34.68&36.32&57.14&98.69&&87.92&100.00&87.92&100.00&100.00&81.97\\
				
				&4&88.21&73.20&64.57&30.71&36.32&57.14&73.96&&74.15&100.00&74.15&100.00&100.00&61.43\\
				
				\hline
				\multirow{4}{*}{Hf\textsubscript{2}COS}&0&99.56&84.41&84.04&38.54&36.32&57.14&100.00&&98.12&98.37&96.52&100.00&100.00&94.80\\
				
				&2&95.59&79.91&76.39&35.44&36.32&57.14&98.36&&87.73&100.00&87.73&100.00&100.00&81.69\\
				
				&4&88.11&73.12&64.43&31.30&36.32&57.14&73.68&&73.99&100.00&73.99&100.00&100.00&61.19\\
				
				&6&79.79&67.99&54.26&27.78&36.32&57.14&52.68&&62.31&100.00&62.31&100.00&100.00&43.75\\
				
				\hline
				Hf\textsubscript{2}COSe&6&100.00&76.40&76.40&29.71&36.32&57.14&98.37&&99.99&87.75&87.74&100.00&100.00&81.70\\
				
				\hline
				\multirow{4}{*}{ZrHfCO\textsubscript{2}}&-1&52.40&30.77&16.13&15.97&36.32&30.95&0.00&&33.73&54.91&18.52&100.00&54.17&0.00\\
				
				&0&47.83&31.04&14.85&14.75&36.32&28.29&0.00&&29.88&57.07&17.05&100.00&49.51&0.00\\
				
				&1&43.81&26.10&11.43&11.37&36.32&21.16&0.00&&26.65&49.27&13.13&100.00&37.04&0.00\\
				
				&3&36.91&17.90&6.61&6.59&36.32&11.10&0.00&&21.44&35.40&7.59&100.00&19.42&0.00\\
				
			\end{tabular}
		\end{ruledtabular}
	\end{table*}
	\subsection{Solar Energy Conversion Efficiency}
	The real assessment of the utility of a photocatalyst for water splitting is incomplete unless its solar energy conversion efficiency is quantified. The relevant parameters are light absorption efficiency, carrier utilisation efficiency, and STH efficiency. Assuming 100\% efficiency of the catalytic reactions, the efficiency of light absorption is defined as \cite{m2x3}
	\begin{eqnarray}
	\eta_{abs}&=&\dfrac{\int^{\infty}_{E_g} P(\hbar \omega) d(\hbar \omega)}{\int^{\infty}_0 P(\hbar \omega) d(\hbar \omega)}
	\end{eqnarray}
	P($\hbar\omega$) denotes the incident AM1.5G solar flux at photon energy $\hbar\omega$, and E$_g$ is the band gap of materials. The quantity in the numerator is the power density absorbed by the catalyst while the one in the denominator stands for the incident power density of sunlight (AM 1.5G), implying that $\eta_{abs}$ is the ratio of converted energy to the total energy. A related quantity is the quantum efficiency of the catalyst, defined as the ratio of the number of utilised photons to the total number of incident photons,
	\begin{eqnarray}
	\eta_{abs}^Q=\dfrac{\int^{\infty}_{E_g} \dfrac{P(\hbar \omega)}{\hbar\omega} d(\hbar \omega)}{\int^{\infty}_0 \dfrac{P(\hbar \omega)}{\hbar\omega} d(\hbar \omega)}
	\end{eqnarray}
	
	The carrier utilisation efficiency is defined as 
	\begin{eqnarray}
	\eta_{cu}=\dfrac{\Delta G \int^{\infty}_{E} \dfrac{P(\hbar \omega)}{\hbar \omega}d(\hbar \omega)}{\int^{\infty}_{E_g} P(\hbar \omega) d(\hbar \omega)}
	\end{eqnarray}
	The related quantity, quantum efficiency of carrier utilisation is given by \cite{m2x3}
	\begin{eqnarray}
	\eta_{cu}^Q=\dfrac{\int^{\infty}_{E} \dfrac{P(\hbar \omega)}{\hbar \omega}d(\hbar \omega)}{\int^{\infty}_{E_g} \dfrac{P(\hbar \omega)}{\hbar \omega} d(\hbar \omega)}
	\end{eqnarray}
	$\Delta$G = 1.23 eV is the energy difference between redox potential H$_2$O/O$_2$ and H$^+$/H$_2$.
	The calculation of efficiency crucially depends on the extra energy required to overcome the barriers of HER and OER, which can be quite large. Co-catalysts are often used to circumvent this difficulty in experiments. In previous experiments, it was reported that the overpotentials of IrO$_x$, NiFeO$_x$, NiCoO$_x$, and CoFeO$_x$ co-catalysts for OER are below 0.5 eV \cite{mccrory2013benchmarking} and that of Pt co-catalysts for HER is lower than 0.1 eV \cite{zheng2015advancing}. Therefore, considering energy loss due to carrier migration between two surfaces of the catalyst, we assume an overpotential of 0.6 eV(0.2 eV)  for OER(HER). Practically, the photons which have energy greater than $E$ are utilised in water-splitting reactions while the materials absorb energy greater than E$_g$. Then, $E$ is determined as follows. 
	\begin{eqnarray}
	E& =& E_g; \chi\left(H_2 \right) \geq 0.2 eV, \chi\left(O_2 \right) \geq 0.6 eV \notag \\
	& =& E_g+0.6-\chi\left(O_2 \right); \chi\left(H_2 \right) \geq 0.2 eV, \chi\left(O_2 \right) < 0.6 eV \nonumber \\
	& =& E_g+0.2-\chi \left(H_2 \right); \chi\left(H_2 \right) < 0.2 eV, \chi\left(O_2 \right) \geq 0.6 eV \nonumber \\
	& =& E_g+0.8-\chi\left(H_2 \right)-\chi\left(O_2 \right); \chi\left(H_2 \right) < 0.2 eV,  \nonumber \\
	& & \enskip \enskip \enskip \enskip \enskip \enskip \enskip \enskip \enskip \enskip \chi\left(O_2 \right) < 0.6 eV
	\label{E}
	\end{eqnarray}
	Clearly, large $\chi\left(H_2 \right)$ and $\chi\left(O_2 \right)$ lead to greater $\eta_{cu}$ as the energy window of photon absorption increases.
	Using the full solar spectrum, the Solar-to-Hydrogen conversion efficiency and associated quantum efficiency are calculated as  \cite{m2x3}: 
	\begin{align}
	& \eta_{STH}=\eta_{abs} \times \eta_{cu}\\%=\dfrac{\Delta G \int^{\propto}_{E} \dfrac{P(\hbar \omega)}{\hbar \omega}d(\hbar \omega)}{\int^{\propto}_0 P(\hbar \omega) d(\hbar \omega)}\\
	& \eta_{T}^Q=\eta_{abs}^Q \times \eta_{cu}^Q%=\dfrac{\int^{\propto}_{E} \dfrac{P(\hbar \omega)}{\hbar \omega}d(\hbar \omega)}{\int^{\propto}_0 \dfrac{P(\hbar \omega)}{\hbar \omega} d(\hbar \omega)}
	\end{align}
	The intrinsic electric fields across the surfaces in M$_{2}$COT compounds help in the separation of electron-hole pairs. The work done by this field is, therefore, to be taken into account. The STH efficiency ($\eta'_{STH}$), after taking this into consideration is \cite{m2x3}:
	\begin{equation}
	\eta'_{STH}=\dfrac{\Delta G \int^{\infty}_{E} \dfrac{P(\hbar \omega)}{\hbar \omega}d(\hbar \omega)}{\int^{\infty}_0 P(\hbar \omega) d(\hbar \omega)+\Delta \Phi\int^{\infty}_{E_g} \dfrac{P(\hbar \omega)}{\hbar \omega}d(\hbar \omega)}
	\end{equation}
	where $\Delta\Phi$ is the potential difference between two surfaces of  Janus Mxene. 
	\begin{table*}[htb!]
		\small
		\caption{Carrier Effective Mass ($m^{*}_i$), In Plane Stiffness Constants ($C_i$), Deformation Potentials ($E_i$) and Carrier Mobilities ($\mu_i$) of electrons and holes along $x$ and $y$ directions of Janus MXenes at their equilibrium volumes. The Carrier Effective Masses are presented in the unit of electron mass $m_0$=9.11$\times$10$^{-31}$ kg.}
		\label{Table3}
		\begin{ruledtabular}
			\begin{tabular}{cccccccccc}
				\textbf{Compounds}&\textbf{Carrier} & \textbf{m$^{*}_x$} &\textbf{m$^{*}_y$} & \textbf{C$_x$}& \textbf{C$_y$}& \textbf{E$_x$}& \textbf{E$_y$}& $\bm{\mu_x}$& $\bm{\mu_y}$\\
				&\textbf{Type}&(m$_0$)&(m$_0$)&(Nm$^{-1}$)&(Nm$^{-1}$)&(eV)&(eV)&(cm$^2$V$^{-1}$s$^{-1}$)&(cm$^2$V$^{-1}$s$^{-1}$)\\
				\hline
				\multirow{2}{*}{Sc\textsubscript{2}COS} & e$^{-}$ &0.935 &0.965&\multirow{2}{*} {123.965}& \multirow{2}{*} {123.783}&3.76 & 3.74&142.91 &135.40  \\
				& h$^{+}$ & 0.829&0.551 & & & 1.43&1.33 &1256.81 &3284.03  \\
				\hline
				\multirow{2}{*}{Zr\textsubscript{2}COS} & e$^{-}$ &2.232 &0.161&\multirow{2}{*} {212.210}& \multirow{2}{*} {212.060}& 3.88&2.58 &40.31 &17511.38  \\
				& h$^{+}$ &0.194 &0.194 & & &4.29 &4.82 &4365.16 &3455.52  \\
				\hline
				\multirow{2}{*}{Hf\textsubscript{2}COS} & e$^{-}$ &1.887 &0.147&\multirow{2}{*}{234.772}& \multirow{2}{*}{234.495}&3.30 &2.86 &46.26 &18902.51  \\
				& h$^{+}$ & 0.392&0.182 & & &5.19 &5.30 &808.15 &3590.80  \\
				\hline
				\multirow{2}{*}{Hf\textsubscript{2}COSe 6\%} & e$^{-}$ &3.197 &0.173&\multirow{2}{*}{142.741}& \multirow{2}{*}{135.349}& 2.82&2.61 &25.02 &9458.77  \\
				& h$^{+}$ & 0.325&0.160 & & & 7.32&7.81 &359.35 &1235.00  \\
				\hline
				\multirow{2}{*}{ZrHfCO\textsubscript{2}} & e$^{-}$ &2.044 &0.275&\multirow{2}{*}{278.055}& \multirow{2}{*}{277.925}&7.98 &8.13 &14.89 &792.20  \\
				& h$^{+}$ & 0.215&0.215 & & &3.54 &3.13 &6839.12 &8744.08  \\
			\end{tabular}
		\end{ruledtabular}
	\end{table*}
	
	In Table \ref{Table2}, we present these parameters calculated at various $\alpha$ for the five Janus compounds. Additionally, we present results of quantum($\eta^{Q}_{UV},\eta^{Q}_{VIS}, \eta^{Q}_{IR}$)  and energy conversion efficiencies ($\eta_{UV},\eta_{VIS},\eta_{IR}$) of UV, visible and IR lights. The range of IR, Visible, and UV light considered for calculating these are <1.65 eV, 1.65-3.10 eV, and > 3.10 eV, respectively. We find that the maximum corrected STH efficiency $\eta^{\prime}_{STH}$ obtained in the cases of unstrained Zr$_2$COS and Hf$_{2}$COS far exceeds the maximum obtained efficiency in a number of 2D photocatalysts. The maximum efficiency in MoSSe is $\sim$  15\% \cite{ju2020janusMo}, in WSSe $\sim$ 8\%-16\%\cite{ju2020janus}, in SiPN $\sim$ 8.37\%, in SiAsP $\sim$ 11\%, in SiSbP $\sim$ 20\%, in SiSbAs $\sim$13\% \cite{zhao2023janus}, in Janus PtSO $\sim$ 7.62\%, in PtSeO $\sim$23\% \cite{shen2021janus}, in Janus In\textsubscript{2}X\textsubscript{2}X' (X and X'= S, Se, and Te) $\sim$10\%-21\% \cite{wang2021two} and in M\textsubscript{2}X\textsubscript{3} $\sim$ 20-32\%\cite{m2x3}, way lower than $\sim$ 30-39 \% obtained in the IR-active systems considered in this work. Among the two visible-active systems, Sc$_{2}$COS has a maximum STH efficiency of $\sim$ 4\% while it is $\sim$ 16 \% for ZrHfCO$_{2}$. This significant dissimilarity in $\eta^{\prime}_{STH}$ of IR-active and visible-active compounds is due to huge differences in energy conversion as well as carrier utilisation efficiencies. 
	
	The origin of such differences can be understood from energy conversion and carrier utilisation efficiencies associated with different parts of the solar spectrum. A look at the quantum efficiencies reveals that the three IR-active MXenes have 100\% UV and visible light absorption efficiencies while it varies between 44-95 \% for IR light. The carrier utilisation efficiency is between 88-100\%. In contrast, the visible-active MXenes have no efficiency in the absorption of IR light. Even the efficiencies are only about 55\%(for ZrHfCO$_{2}$) for visible light. In Sc$_{2}$COS, it is remarkably poor ($\sim$ 1-11 \%). The carrier utilisation efficiencies vary between 19-42 \% in Sc$_{2}$COS and between 35-55 \% in ZrHfCO$_{2}$. The large band gaps in these two compounds lead to absorption of only 6-12 \% of photons for Sc$_{2}$COS while it is 21-33 \% for ZrHfCO$_{2}$. In comparison, 62-100 \% of photons are absorbed by the IR-active compounds. Strained Hf$_{2}$COSe, the one with the best absorbance in the IR region, absorbs almost 100 \% photons. But its carrier utilisation efficiency is only about 88\%, not as great as the other two IR-active MXenes. The energy conversion efficiency of this compound, however, is about $\sim$ 30 \%, close to Zr$_{2}$COS and Hf$_{2}$COS at 4\% strain. 
	
	The overpotentials that are very important in shaping the efficiency parameters of a photocatalyst can be understood from their variations with strain. Hf$_{2}$COSe having the smallest band gap of 0.2 eV (Table \ref{Table1}) has the largest $\eta_{abs}$ but its $\eta_{cu}$ is less than that of Zr$_{2}$COS and Hf$_{2}$COS when the later two are subjected to a tensile strain upto 2\%. The origin of this lies in the value of $E$, the minimum energy of photons practically used for water splitting, and the size of the bang gap (Equation (\ref{E})). Though Hf$_{2}$COSe has large $\chi\left(H_2 \right)$, only 0.27 eV $\chi\left(O_2 \right)$ implies $E=0.53$ eV, higher than 0.51(0.47) eV of Zr$_{2}$COS(Hf$_{2}$COS) at 0 \% strain. At 2\% strain, although Zr$_2$COS(Hf$_{2}$COS) is 0.63eV(0.66eV) greater than that of Hf$_{2}$COSe, the much smaller E$_g$ of Hf$_{2}$COSe makes the denominator in Equation (5) dominating, resulting in lower  $\eta_{cu}$ of it compared to that of Zr$_2$COS and Hf$_{2}$COS. Substantial increase in their $\chi\left(H_2 \right)$ and $E_{g}$ as tensile strain goes beyond 2\% implies Zr$_{2}$COS(Hf$_{2}$COS) can use photons with energy 0.86(0.9) eV and above. This reduces their $\eta_{cu}$ at higher strain. The variations in $\eta_{abs}$ and $\eta_{cu}$ with strain in Sc$_{2}$COS and ZrHfCO$_{2}$ can be explained in a similar way.
	\subsection{Carrier Mobility}
	Carrier mobility is another crucial parameter in deciding the efficiency of a photocatalyst. In this work, we obtain the carrier mobility using the Deformation Potential (DP) theory \cite{bardeen1950deformation}. In this theory, carrier mobility is determined primarily by the interactions of electrons and acoustic phonons. The carrier mobility along direction $i$ is calculated using the expression 
	\begin{equation}
	\mu_i=\dfrac{2e\hbar^3C_i}{3k_BT|m_i^*|^2E_i^2}
	\end{equation}
	Here, e, $k_B$, and T  represent the electronic charge, Boltzmann constant, and temperature, respectively. $m_i^*= \hbar^2\left[ \dfrac{\partial^2E_k}{\partial k^2}\right]^{-1/2}$ is the carrier effective mass derived from fitting the band edges where  $E_k$ is the band energy corresponding to wave vector $k$. $E_i=\left( \dfrac{\partial E_{edge}}{\partial\delta_i}\right) $ is the Deformation Potential that denotes the shift of band edges $E_{edge}$ (CBM/VBM) under uniaxial strain $\delta_{i}$ along direction $i$. The in-plane stiffness constant along direction $i$ is determined by $C_i=\dfrac{1}{S_0}\left( \dfrac{\partial^2E_{total}}{\partial \delta_i^2}\right) $ where E$_{total}$ is the total energy of the system and S$_0$ is the surface area of the equilibrium volume. $C_i$ and $E_i$ are obtained by varying $\delta_i$ from -2\%to 2\% along $x$ and $y$ directions and fitting the corresponding changes in $E_{total}$ and $E_{edge}$ to appropriate functions. Calculated values of stiffness constants ($C_i$), effective masses ($m_i^*$), deformation potentials ($E_i$), and mobilities ($\mu_i$) of each carrier type are shown in Table \ref{Table3}. The results are for systems without strain except for Hf$_{2}$COSe, where the system under 6\% tensile strain is considered.
	
	The results clearly show qualitative differences between carrier mobilities of IR-active and visible-active systems. For visible-active Sc$_{2}$COS and ZrHfCO$_{2}$, electron mobility, irrespective of direction, is the order of magnitude less than the hole mobility. For IR-active Zr$_{2}$COS, Hf$_{2}$COS, and Hf$_{2}$COSe, the relative strengths of carrier mobility is direction dependent: holes(electrons) have an order of magnitude greater mobility along $x(y)$-direction. However, the common feature in all of them is the significant difference in electron and hole carrier mobilities. Larger effective mass and deformation potential of electrons are responsible for significant small electron mobility in Sc$_{2}$COS and ZrHfCO$_{2}$. These, in turn, occur due to the relatively flat CBM of these compounds. In the three IR-active MXenes, the electron and hole effective masses along $y$ are nearly the same. Along $x$-direction, electron effective masses are the order of magnitude larger. This explains why hole mobility along $x$ is orders of magnitude larger in these compounds. On the other hand, the reason behind the order of magnitude larger electron mobility with respect to that of holes along $y$ is driven by the larger deformation potential of holes with respect to that of electrons. 
	
	The separation of carriers with opposite polarity enhances, resulting in less probability of carrier recombination if mobilities of different carriers have different preferential directions. For visible-active Sc$_{2}$COS(ZrHfCO$_{2}$), electrons(holes) have no preferential direction for migration while holes(electrons) have greater mobility along $y$. Among the three IR-active photocatalysts, both carriers in Hf$_2$COT (T= S, Se) have greater preference along $y$. Different carriers in Zr$_{2}$COS have different preferential directions for migration. While it is $y$ for electrons, holes have greater preference along $x$. Thus, for all five compounds, good carrier separation leading to less probability of carrier recombination is prominent.
	
	Upon comparing the carrier mobilities of well known 2D photocatalysts ( 270 cm$^2$V$^{-1}$s$^{-1}$, 130 cm$^2$V$^{-1}$s$^{-1}$ for MoS\textsubscript{2}, 90 cm$^2$V$^{-1}$s$^{-1}$, 25 cm$^2$V$^{-1}$s$^{-1}$ for MoSe\textsubscript{2}, 540 cm$^2$V$^{-1}$s$^{-1}$, 320 cm$^2$V$^{-1}$s$^{-1}$ for WS\textsubscript{2}, 270 cm$^2$V$^{-1}$s$^{-1}$, 30 cm$^2$V$^{-1}$s$^{-1}$ for WSe\textsubscript{2}\cite{jin2014intrinsic}, 210 cm$^2$V$^{-1}$s$^{-1}$, 53 cm$^2$V$^{-1}$s$^{-1}$ for MoSSe \cite{ju2020janusMo}, 723 cm$^2$V$^{-1}$s$^{-1}$, 125 cm$^2$V$^{-1}$s$^{-1}$ for WSSe \cite{ju2020janus} ) we find that carrier mobilities in Janus compounds considered here are far better. However, a comparison of mobilities among these Janus and their endpoint compounds M$_{2}$CO$_{2}$  produces mixed results. Mobilities of Hf$_{2}$COT are far greater than Hf$_{2}$CO$_{2}$ where $\mu^h\sim$ 1.6 to 2.2 $\times$10$^3$ cm$^2$V$^{-1}$s$^{-1}$, $\mu^e\sim$ 0.12 to 2.2 $\times$10$^3$ cm$^2$V$^{-1}$s$^{-1}$ \cite{guo2016mxene}. Same happens upon comparison between Zr$_{2}$COS and Zr$_{2}$CO$_{2}$. The later has hole(electron) mobilities of 2.3$\times$10$^3$ cm$^2$V$^{-1}$s$^{-1}$ and 1.7$\times$10$^3$ cm$^2$V$^{-1}$s$^{-1}$(83 cm$^2$V$^{-1}$s$^{-1}$ and 1.1$\times$10$^3$ cm$^2$V$^{-1}$s$^{-1}$) along two directions. In case of ZrHfCO$_{2}$, the hole(electron) mobilities are higher(lower) than end point compounds Zr$_{2}$CO$_{2}$ and Hf$_{2}$CO$_{2}$ while the mobilities of both carriers in Sc$_{2}$COS are significantly lower than the conventional counterpart Sc$_{2}$CO$_{2}$ ($\mu^h\sim$ 6.4 to 9.5 $\times$10$^3$ cm$^2$V$^{-1}$s$^{-1}$, $\mu^e\sim$ 6.1 to 12.5 $\times$10$^3$ cm$^2$V$^{-1}$s$^{-1}$ \cite{wang2023first}).  
	\begin{table*}[htb!]
		\small
		\caption{Variations in the Carrier Effective Mass ($m^*_{ex}, m^*_{hx}, m^*_{ey}, m^*_{hy}$), Reduced Mass ($\mu_{rx},\mu_{ry}$) along $x$ and $y$ directions, Static Dielectric Constant ($\varepsilon(0)$), 2D Polarisability ($\chi_{2D}$), theoretically predicted and numerically calculated Exciton Extensions ($\lambda_{th}, \lambda$), Exciton Bohr radii ($a_x, a_y$) and Exciton Binding Energy ($E_b$) of Janus MXenes with Biaxial strain ($\alpha$), calculated using the Hydrogenic Exciton model.}
		\label{Table4}
		\begin{ruledtabular}
			\begin{tabular}{ccccccccccccccc}
				\textbf{Compounds}&\textbf{$\bm{\alpha}$}&$\bm{m^*_{ex}$}&$\bm{m^*_{hx}$}&$\bm{\mu_{rx}}$&$\bm{m^*_{ey}$}&$\bm{m^*_{hy}$}&$\bm{\mu_{ry}}$&$\bm{\varepsilon(0)} $&$\bm{\chi_{2D}}$&$\bm{\lambda_{th}}$&$\bm{\lambda}$&$\bm{a_x$}&$\bm{a_y$}&$\bm{E_b$  }\\
				&(\%)&(m$_0$)&(m$_0$)&(m$_0$)&(m$_0$)&(m$_0$)&(m$_0$)&&(\AA)&\resizebox*{0.085\textwidth}{!}{$\sim(\dfrac{\mu_{rx}}{\mu_{ry}})^{1/3}$}&&(\AA)&(\AA)&(eV)\\
				\hline
				\multirow{5}{*}{Sc\textsubscript{2}COS}&-4&1.172&0.813&0.48&1.168&0.553&0.38&2.65&2.63&1.09&1.09&4.82&5.24&1.23\\
				
				&-2&0.988&0.819&0.45&1.003&0.558&0.36&2.67&2.65&1.08&1.08&5.01&5.41&1.21\\
				
				&0&0.935&0.829&0.44&0.965&0.551&0.35&2.70&2.71&1.08&1.08&5.11&5.52&1.18\\
				
				&2&0.948&0.833&0.44&0.997&0.542&0.35&2.75&2.78&1.08&1.08&5.15&5.58&1.16\\
				
				&4&0.985&0.838&0.45&1.065&0.529&0.35&2.80&2.87&1.09&1.09&5.17&5.63&1.14\\
				\hline
				\multirow{3}{*}{Zr\textsubscript{2}COS}&0&2.232&0.194&0.18&0.161&0.194&0.09&6.47&8.70&1.27&1.28&14.97&19.09&0.37\\
				
				&2&2.592&0.361&0.32&0.191&0.361&0.12&5.57&7.27&1.36&1.38&10.35&14.24&0.48\\
				
				&4&3.196&0.358&0.32&0.223&0.358&0.14&5.17&6.64&1.33&1.34&9.76&13.06&0.52\\
				
				\hline
				\multirow{4}{*}{Hf\textsubscript{2}COS}&0&1.887&0.392&0.32&0.147&0.182&0.08&5.93&7.85&1.59&1.61&11.17&17.95&0.42\\
				
				&2&2.061&0.385&0.32&0.170&0.184&0.09&5.12&6.56&1.54&1.56&10.19&15.94&0.48\\
				
				&4&2.306&0.380&0.33&0.193&0.185&0.09&4.75&5.97&1.51&1.53&9.66&14.80&0.52\\
				
				&6&2.656&0.375&0.33&0.218&0.190&0.10&4.56&5.67&1.48&1.50&9.33&13.97&0.55\\
				
				\hline
				Hf\textsubscript{2}COSe&6&3.197&0.325&0.30&0.173&0.160&0.08&6.91&9.41&1.53&1.54&12.59&19.44&0.37\\
				
				\hline
				\multirow{4}{*}{ZrHfCO\textsubscript{2}}&-1&1.955&0.214&0.19&0.253&0.214&0.12&3.57&4.09&1.18&1.19&10.00&11.93&0.66\\
				
				&0&2.044&0.215&0.19&0.275&0.215&0.12&3.51&4.00&1.17&1.18&9.83&11.59&0.67\\
				
				&1&2.152&0.514&0.41&0.295&0.217&0.13&3.48&3.94&1.49&1.51&6.99&10.56&0.76\\
				
				&3&2.419&0.498&0.41&0.348&0.217&0.13&3.45&3.89&1.46&1.47&6.90&10.18&0.78\\
			\end{tabular}
		\end{ruledtabular}
	\end{table*}
	\begin{figure*}[htb!]
		\includegraphics[scale=0.44]{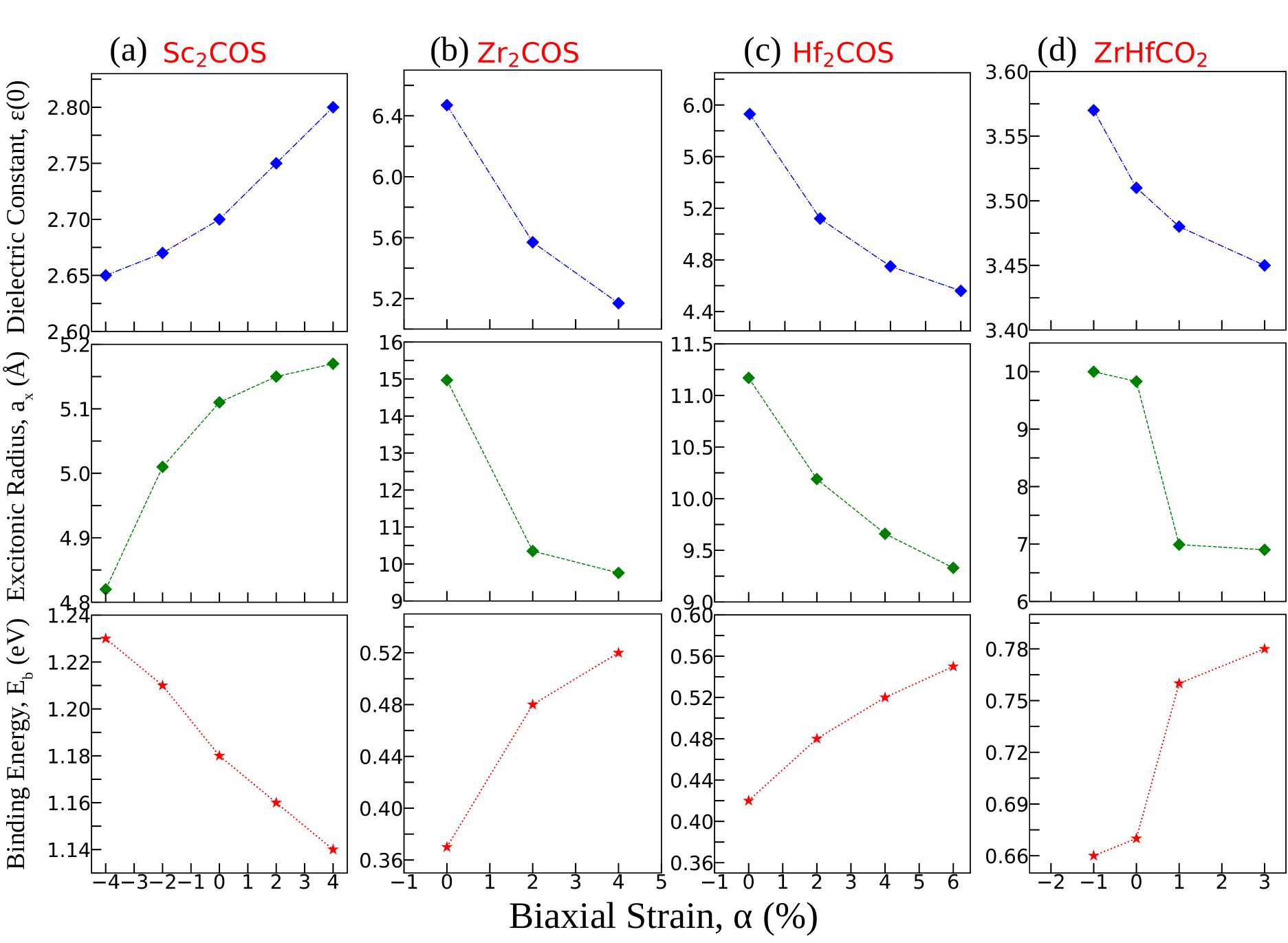}
		\caption{Variations of Static Dielectric Constant $\varepsilon(0)$, Exciton Radius ($a_x$) and Exciton Binding Energy ($E_b$) of a) Sc\textsubscript{2}COS, b) Zr\textsubscript{2}COS, c) Hf\textsubscript{2}COS and d) ZrHfCO\textsubscript{2} with Biaxial strain ($\alpha$).}
		\label{Figure4}
	\end{figure*}
	\subsection{Exciton Binding Energy}
	The Coulomb interaction between the photo-excited electron and hole pairs is attractive, resulting in exciton-bound states. The exciton binding energy representing the strength of this interaction directly affects the photocatalytic efficiency. Excitons can transfer momentum and energy but no charge to the radicals, reducing the efficiency of the water-splitting process. In 2D materials, the reduced dielectric screening enhances the exciton binding energy. A first-principles-based investigation in this regard requires the coupling of many-body perturbation theory to the Bethe-Salpeter equation. This approach is computationally expensive. The difficulty can be circumvented by adopting a 2D hydrogenic exciton model for an anisotropic system within the Keldysh formalism \cite{prada,cudazzo,arra2019}. Several comparisons with the many-body perturbation theory-based GW-BSE method have demonstrated the accuracy of this model\cite{arra,berkel,xu2017}. Therefore, in this work, we have used this model to calculate the exciton binding energies of the Janus MXenes. In this model, the Hamiltonian for the excitons is 
	\begin{equation}
	H=-\dfrac{\hbar^2}{2}\left( \dfrac{1}{\mu_{rx}}\dfrac{\partial^2}{\partial x^2}+\dfrac{1}{\mu_{ry}}\dfrac{\partial^2}{\partial y^2}\right) +V_{2D}(x,y)
	\end{equation}
	Here, $\mu_{rx}$ \& $\mu_{ry}$ are the reduced masses of the excitons along $x$ and $y$ directions and can be derived by the following relation: $\mu_{ri}$= $\left( m_{ei}^{*-1}+m_{hi}^{*-1}\right)$, $i$= x, y. $m_{ei},m_{hi}$ are the effective mass of electrons and holes along $i$, respectively. The Coulomb potential between electron and hole in a 2D dielectric substrate can be represented as 
	\begin{align}
	V_{2D}(r)=-\dfrac{e^2}{4(\varepsilon_1+\varepsilon_2)\varepsilon_0r_0}\left[ H_0\left( \dfrac{r}{r_0}\right)-Y_0\left( \dfrac{r}{r_0}\right) \right] 
	\end{align}
	Here, $H_0$ and $Y_0$ are the zeroth order Struve and Bessel functions, respectively. $\varepsilon_{1,2}$ are the dielectric constants of the two mediums (both vacuum in our case), and $\varepsilon_0$ denotes the vacuum permittivity. $r_0$, the screening length is evaluated using 2D polarizability $\chi_{2D}$ :
	$r_0$=$\left( \dfrac{4\pi}{\varepsilon_1+\varepsilon_2}\right) \chi_{2D}$. $\chi_{2D}$ is related to the static dielectric constant $\varepsilon$(0) of the 2D material : $\varepsilon$(0)=$\left( 1+\dfrac{4\pi\chi_{2D}}{L}\right) $, $L$ is the size of the vacuum between two mono-layers(20 \mbox{\AA} in our case). For $r>r_0$, the potential behaves like a simple Coulomb potential while for $r<r_0$, it diverges logarithmically and can be expressed as \cite{cudazzo}:
	\begin{align}
	V_{2D}^C(r)=-\dfrac{e^2}{4\pi\tilde{\varepsilon}\varepsilon_0r_0}\left[ ln\left( \dfrac{r}{r+r_0}\right) +[\gamma-ln(2)]e^{-\left( \dfrac{r}{r_0}\right) }\right] 
	\end{align}
	$\gamma\approx$ 0.5772 is the Euler constant and $\tilde{\varepsilon}=\dfrac{\varepsilon_1+\varepsilon_2}{2}$.
	
	Using the variational method, the exciton wave function for an anisotropic material like Janus MXenes can be written as : 
	\begin{align}
	\Psi(x,y) & =\sqrt{\dfrac{2}{\pi a_xa_y}} exp\left[ -\left\lbrace\left( \dfrac{x}{a_x}\right)^2 +  \left( \dfrac{y}{a_y}\right)^2 \right\rbrace ^{1/2}\right] \notag \\
	& =\sqrt{\dfrac{2}{\pi \lambda a_x^2}} exp\left[ -\left\lbrace\left( \dfrac{x}{a_x}\right)^2 +  \left( \dfrac{y}{\lambda a_x}\right)^2 \right\rbrace ^{1/2}\right]
	\end{align}
	Here, a$_x$, a$_y$=$\lambda a_x$ are the excitonic radii along the x and y direction, and $\lambda$ is the variational anisotropy scaling factor. Using the above trial wave function,
	the variational exciton binding energy is obtained as
	\begin{equation}
	E_b^{2D}(a_x, \lambda)=E_{kin}^{2D}(a_x, \lambda)+E_{pot}^{2D}(a_x, \lambda).
	\label{tot}
	\end{equation}
	where
	\begin{align}
	E_{kin}^{2D}(a_x, \lambda) & = -\dfrac{\hbar^2}{2} \iint \Psi^*\left[ \dfrac{1}{\mu_{rx}}\dfrac{\partial^2\Psi}{\partial x^2}+\dfrac{1}{\mu_{ry}}\dfrac{\partial^2\Psi}{\partial y^2}\right] dx dy \notag \\
	& =\dfrac{\hbar^2}{4a_x^2}\left[ \dfrac{1}{\mu_{rx}}+\dfrac{1}{\lambda^2\mu_{rx}}\right] 
	\label{kin}
	\end{align}
	and
	\begin{equation}
	E_{pot}^{2D}(a_x, \lambda)=\iint V_{2D}^C(x,y)|\Psi(x,y)|^2 dx dy
	\label{pot}
	\end{equation}
	are the contributions from kinetic and potential energies, respectively.
	The  equations(\ref{tot})-(\ref{pot}) are solved numerically. E$_b^{2D}$($a_x, \lambda$) is minimised with respect to the variational parameters $a_x$ and  $\lambda$ to obtain the actual binding energy E$_b$.
	
	The computed values of the carrier effective masses ($m^{*}_{ei},m^{*}_{hi}$), static Dielectric constant $\varepsilon (0)$, exciton radii (a$_x$, a$_y$), and exciton binding energy ($E_{b}$) of the five Janus MXenes with variation of strain $\alpha$ are presented in Table.\ref{Table4}. Variations of the last three quantities with strain are also plotted in Figure \ref{Figure4}. Comparison between numerically obtained ($\lambda$) and theoretically calculated ($\lambda_{th} \sim(\dfrac{\mu_{rx}}{\mu_{ry}})^{1/3}$ \cite{prada}) anisotropy parameter is also presented in Table \ref{Table4}. We find excellent agreement between these two for all systems and at all values of $\alpha$. We find lower $E_{b}$ for the IR-active photocatalysts in comparison to the visible-active ones with Sc$_{2}$COS producing the highest values.   
	
	The origin of the qualitative behavior of $E_{b}$ across compounds and strain can be understood from the behavior of static dielectric constant and exciton radii. Larger $\epsilon\left(0 \right)$ amounts to larger dielectric screening between electron-hole pairs. This is reflected in larger screening length $r_{0}$ and exciton radii $a_{i}$. From Table \ref{Table4}, we find that the static dielectric constant $\epsilon \left(0 \right)$ is much larger in the three IR-active Janus MXenes in comparison with the visible-active ones. As a result, these systems have larger polarisability $\chi_{2D}$. The screening length $r_{0}$ and exciton radii $a_{i}$ are also larger as a consequence affecting lower $E_{b}$. Among the two visible-active photocatalysts, lower $\epsilon\left(0 \right)$ and $\chi_{2D}$ in Sc$_{2}$COS results in lower screening length, lower exciton radii and higher $E_{b}$ as a consequence. With strain $E_{b}$ decreases(increases) marginally in Sc$_{2}$COS(ZrHfCO$_{2}$) as a consequence of directions of variations in $\epsilon \left(0 \right), \chi_{2D}, r_{0}$ and $a_{i}$. The same happens for IR-active Zr$_{2}$COS and Hf$_{2}$COS. With increasing strain, the opposite trends in Sc$_{2}$COS and the rest can be traced back to the variations in the electronic band gaps. With increasing strain, $E_{g}$ decreases in Sc$_{2}$COS while it increases for the rest. The decrease in $E_{g}$ leads to larger dielectric screening. In fact, the root of the qualitative trend of $E_{b}$ across compounds can be explained this way.
	
	The exciton binding energies of the Janus compounds considered in this work compare very well with those of well-known 2D photocatalysts. The range of $E_{b}$ in our case, 0.37-1.23 eV, is in good agreement with MoS\textsubscript{2} ($E_{b}$=1.03 eV) \cite{ramasubramaniam2012large}, MoSe\textsubscript{2} ($E_{b}$=0.91 eV) \cite{ramasubramaniam2012large}, WS\textsubscript{2} ($E_{b}$=1.04 eV) \cite{ramasubramaniam2012large}, WSe\textsubscript{2} ($E_{b}$=0.90 eV) \cite{ramasubramaniam2012large}, WSSe ($E_{b}$=0.83 eV)  \cite{ju2020janus}, GeS ($E_{b}$=0.98 eV) \cite{xu2017}, GeSe ($E_{b}$=0.38 eV) \cite{xu2017}, GeTe ($E_{b}$=0.21 eV) \cite{xu2017}, SnS ($E_{b}$=0.50 eV) \cite{xu2017} and SnSe ($E_{b}$=0.28 eV) \cite{xu2017}.
	
	\section{Conclusions}
	Extensive DFT-based calculations have been performed on 14 compounds from the family of M$_{2}$COT (M=Sc, Ti, Zr, Hf; T=S, Se) and MM$^{\prime}$CO$_{2}$(M, M$^{\prime}$=Sc, Ti, Zr, Hf) Janus MXenes to find new photocatalysts for water splitting reactions, active in visible or IR part of the solar spectrum. The compounds have been chosen from a previous elaborate study \cite{swati1} that predicted four new photocatalysts. Bi-axial strain, both compressive and tensile, have been applied on the compounds to either open an electronic band gap in semi-metals and/or to enforce alignments of band edges so that water-splitting reactions are possible. We found that the four existing Janus MXenes that had the proper band alignment to effect water splitting reactions preserved it when subjected to tensile strain up to 4\%. In the course of the study, Hf$_{2}$COSe is found to be a new photocatalyst active in the IR region when subjected to 6\% strain. We compute various performance parameters of these five compounds, varying the amount of tensile strain. We find that IR active Zr$_{2}$COS, Hf$_{2}$COS at equilibrium and low strain along with 6\% strained Hf$_{2}$COSe have absorbance, carrier utilisation, carrier separation, exciton binding, and solar-to-hydrogen conversion efficiency that are far superior than the well known 2D photocatalysts for water splitting reactions. Among the visible-active photocatalysts, Sc$_{2}$COS cannot meet the benchmark of 10\% solar-to-hydrogen conversion efficiency while the performance parameters of ZrHfCO$_{2}$ are, at best, moderate. To our knowledge, this is the first comprehensive work demonstrating the utility of Janus MXenes as catalysts in photo-assisted water-splitting reactions. The work paves the way for further exploration into the family of MXenes to discover more photocatalysts for water-splitting reactions, a sustainable source of green energy. 
	\section*{Acknowledgement}
	The authors gratefully acknowledge the Department of Science and Technology, India, for the computational facilities under Grant No. SR/FST/P-II/020/2009 and IIT Guwahati for the PARAM supercomputing facility where all computations are performed. The authors would like to thank Dr. Mukul Kabir for useful discussions.
%	\twocolumngrid
%	\bibliographystyle{apsrev4-2.bst}
%	\bibliography{references_2.bib}

%apsrev4-2.bst 2019-01-14 (MD) hand-edited version of apsrev4-1.bst
%Control: key (0)
%Control: author (72) initials jnrlst
%Control: editor formatted (1) identically to author
%Control: production of article title (-1) disabled
%Control: page (0) single
%Control: year (1) truncated
%Control: production of eprint (0) enabled
%
\end{document}